\documentclass[conference]{IEEEtran}
\usepackage{listings}
\usepackage{color}
\usepackage[T1]{fontenc}
\usepackage[latin9]{inputenc}
\usepackage[english]{babel}
\usepackage{url}
\usepackage{graphicx}

\usepackage[belowskip=3pt,aboveskip=3pt]{caption}
\usepackage{graphics}
\usepackage{textcomp}
\definecolor{sh_comment}{rgb}{0.12, 0.7, 0.18}
\definecolor{dkgreen}{rgb}{0,0.6,0}
\definecolor{gray}{rgb}{0.5,0.5,0.5}
\definecolor{mauve}{rgb}{0.58,0,0.82}
\definecolor{light-gray}{gray}{0.80}

\usepackage{balance}

\usepackage[font={footnotesize}]{caption}

\begin{document}
\title{Security Toolbox for Detecting Novel and Sophisticated Android Malware$^*$}


\author{
\IEEEauthorblockN{Benjamin Holland, Tom Deering,  Suresh Kothari}
\IEEEauthorblockA{Department of Electrical and Computer Engineering\\
Iowa State University,
Ames, Iowa 50010, Email: \{bholland, tdeering, kothari\}@iastate.edu}
\IEEEauthorblockN{Jon Mathews, Nikhil Ranade}
\IEEEauthorblockA{EnSoft Corp. Email: \{jmathews, nikhil\}@ensoftcorp.com}
}


\maketitle

\begin{abstract}
This paper presents a demo of our Security Toolbox to detect novel malware in Android apps. This Toolbox is developed through our recent research project funded by the DARPA Automated Program Analysis for Cybersecurity (APAC) project. 
The adversarial challenge ("Red") teams in the DARPA APAC program are tasked with designing sophisticated malware to test the bounds of malware detection technology being developed by the research and development ("Blue") teams. Our research group, a Blue team in the DARPA APAC program, proposed a ``human-in-the-loop program analysis'' approach to detect  malware  given the source or Java bytecode for an Android app.
Our malware detection apparatus consists of two components: a general-purpose program analysis platform called Atlas, and a Security Toolbox built on the Atlas platform.
This paper describes the major design goals, the Toolbox  components to achieve the goals, and the workflow for auditing Android apps. The accompanying video illustrates features of the Toolbox through a live audit.  \newline \newline
\vspace{.5 em}
\textbf{Video: http://youtu.be/WhcoAX3HiNU}
\end{abstract}

{\let\thefootnote\relax\footnotetext{*This material is based on research sponsored by DARPA under agreement number FA8750-12-2-0126. The U.S. Government is authorized to reproduce and distribute reprints for Governmental purposes notwithstanding any copyright notation thereon.}}

%\keywords{cybersecurity, Android, mobile apps, malware detection, program analysis}

% \fi
\IEEEpeerreviewmaketitle

\section{Introduction}

Searching for novel malware can be like looking for a needle in the haystack, but without knowing what a needle is or having ever seen one.  In 2010 we learned of Stuxnet \cite{STUXNET}, a targeted nation-state level attack against an Iranian nuclear research site.  The attack was only detected, some speculate intentionally, when it began to utilize noisy traditional attack vectors such as USB malware propagation.  Recently we have seen a proliferation of high-level logic bugs in SSL \cite{Heartbleed, GOTO-Fail} and even a recently discovered 25-year-old logic bug in the Bash shell \cite{Shellshock}.  While most would agree that these bugs were honest mistakes, a few have speculated that some may have been added with malicious intent \cite{Heartbleed-conspiracy}.  Since we have no way to determine intent by examining code a security analyst must consider software bugs as potential malice.  In either case the consequences can be catastrophic. When the stakes are high, the current practices for malware detection are far from adequate. The DARPA APAC program aims at creating new techniques and tools to detect sophisticated Android malware capable of causing serious damage in a Department of Defense scenario.

USAF Colonel John Boyd described the \textbf{OODA loop} as an iterative decision cycle of \emph{observe}, \emph{orient}, \emph{decide}, and \emph{act}. Boyd developed this framework as a way to explain the unanticipated, superior agility of US fighter pilots in aerial combat situations. The paradigm of OODA loops applies equally well to the APAC context. To detect malware, our tools must be able to outmaneuver the capabilities of  adversaries who will continue to develop new varieties of Android malware.  Our Security Toolbox for Android is designed to utilize best-in-class automation and iteration techniques to maximize the odds of emerging victorious from this confrontation. We completed Phase I of the DARPA APAC program as the top performing Blue team.

\section{Design Goals}

\subsection{Minimizing Human Effort}
\noindent \textbf{Goal:} Minimize the human effort for (a) cross-verifying automatically detected malware, (b) performing what-if experiments to hypothesize, refine, and postulate application-specific malware that is not on the radar of automated malware detection. 

We incorporate a Query-Model-Refine (QMR) program analysis platform, called Atlas~\cite{Atlas-paper, ICSE-Demo}, developed by EnSoft; it provides the tool mechanics necessary for our human-in-the-loop detection of malware. 
We use a heterogeneous, attributed, directed graph data structure as an abstraction to represent the essential aspects of the program's syntax and semantics (structure, control flow, and data flow), which are required to reason about software. Atlas constructs this graph from a set of software projects provided by the user. Atlas offers an expressive query language for users to write composable analyzers. Analyzers compute results in the form of subgraphs relevant to the query (evidence), which can be visualized. Based on the evidence, users can issue further queries, possibly involving information beyond specific program artifacts (e.g., looking for a specific URL). The above iteration continues until the user is satisfied with the analysis. The Security Toolbox includes analyzers using the Atlas query language. These analyzers incorporate Android semantics and they can be invoked programmatically or through interactive ``Smart Views,'' described in Section \ref{sec:SmartViews}.

\subsection{Incorporating Android Semantics}
\noindent \textbf{Goal:}  Incorporate rich and complex semantics that Android provides to facilitate development of mobile apps. 

To address the semantics of Android, the Security Toolbox incorporates the permission mapping between Android APIs and the permissions each API requires. The Toolbox also incorporates semantics of fundamental Android components such as
Activities, Services, Content providers, and Broadcast receivers\footnote{\url{https://developer.android.com/guide/components/fundamentals.html}}, as well as Android specific XML resources\footnote{\url{https://developer.android.com/guide/topics/resources/providing-resources.html}}.

We have developed new algorithms to automatically summarize all the Android APIs. This is work in progress and when completed we will incorporate these summaries in our Toolbox and also make them available to others in a portable format.  

\subsection{Evolution and User-friendly Design}
\noindent \textbf{Goal:}  Develop a detection tool that is evolution-friendly and highly usable.

We have a decoupled architecture to achieve this goal. The malware detection capability is decoupled and built on top of the program analysis platform (Atlas). 
The underlying design philosophy is similar to platforms like Matlab or Mathematica with domain-specific toolboxes built on top of general-purpose machinery.  
The low-level static analysis resides inside Atlas, and the malware detection capability resides inside the Toolbox as analyzers using Atlas queries. Refining and extending the existing detection capabilities as well as creating entirely new capabilities is relatively easy because it can all be done through query-enabled analyzers. Since creating a complete list of malware properties is unrealistic, it is imperative that it be relatively simple to expand the cookbook of ready-made properties through the use of adversarial thinking.
\section{Use Cases for the Security Toolbox}

The Toolbox is useful for nearly any Android malware detection task, with three main use cases described as follows:
\begin{itemize}
\item Automated detection of Android malware that has a clearly defined specification
\item Production of evidence to support conclusions of automated analysis
\item Enabling the human to perform what-if experiments to hypothesize and detect new malware that cannot be detected automatically because its pattern or specification is not known a priori 
\end{itemize}

\section{Components}
\label{sec:Components}
The Security Toolbox is logically separated into several components as detailed below.  

\subsection{Permission Mapping}
Android's sensitive functionalities such as sending and receiving text messages, accessing geo-location information, or accessing user contacts are protected by runtime checks that enforce whether or not an application has been granted permission to invoke such functionalities. 
The Security Toolbox leverages the permission mapping produced by the Toronto PScout research group \cite{Au:2012:PAA:2382196.2382222}.  For each API version of Android, we transform the PScout mapping to an XML file that precisely represents the permission protected methods.  The Toolbox contains code for parsing an Application's manifest, and uses the XML file to automatically annotate the correct API mapping onto the Atlas program graph.  We have automatically scraped and encoded into Java objects the Google developer documentation for permissions, permission groups, and protection levels to aid in developing analyzers.  Additionally we have recovered mappings for Android permissions to protection levels, and permissions to permission groups by mining their relationships from the Android source\footnote{\url{https://github.com/EnSoftCorp/android-essentials-toolbox}}. 

\subsection{Indexers}
The Atlas program graph provides much of the information needed to analyze programs, but some information is a conservative estimate.  One example is type inference where the dynamic dispatch edges may be conservatively resolved to many potential targets.  To address this problem, the Security Toolbox implements a Rapid Type Analysis (RTA) \cite{RTATypeAnalysis} strategy to exclude call edges to methods that should not be possible at runtime based on observed constructor calls.  The Type Inference Indexer performs the RTA analysis and annotates the edges in the program graph for use by other analyzers.  
Since Android makes extensive use of XML for its user interface, manifest, and other resources many important program artifacts are missing in the Java program graph produced by Atlas.  The Security Toolbox provides indexers to annotate and add missing program elements from these resources to the Atlas program graph.

\subsection{Analyzers}
The Security Toolbox defines an Analyzer Interface that encapsulates the logic for traversing a program graph to extract an "envelope" (a subgraph that is either empty if the security property is satisfied or non-empty containing the necessary information to locate the violation of the security property).  Analyzers encapsulate their descriptions, assumptions, and possible continuations to refine results or broaden a traversal.  For example, one possible continuation for a data flow based taint analysis between a sensitive source and a sensitive sink that produced a graph that is too large to interpret would be to perform the same taint analysis with call, object, type, and flow sensitivities enabled.  Analyzers have been subdivided into property, smell, confidentiality, integrity, and availability analyzers.  A property is something the analyst should be aware of, but does not necessarily indicate malice, such as uses of native code.  A smell is a heuristic similar to a property that indicates a stronger suspicion, which demands a justification such as using Java reflection to invoke a private API.  The confidentiality, integrity, and availability (CIA) analyzers detect violations of CIA properties using taint analysis of sources and sinks, modification operations on sensitive mutables, and loop detection of expensive resources respectively.

\lstset{language=Java,
numbers=left,
numbersep=5pt,
backgroundcolor=\color{white},
showspaces=false,
showstringspaces=false,
showtabs=false,
frame=single,
tabsize=2,
captionpos=b,
breaklines=true,
breakatwhitespace=false,
frame=shadowbox,
rulesepcolor=\color{blue},
keywordstyle=\bfseries\color{blue},
morekeywords={public, Malware, trigger, block, payload, class, extends,
protected, void, if, new},
escapechar=!,
%stringstyle=\color{red},
commentstyle=\color{sh_comment},
%basicstyle=\scriptsize,
frame=tb,
%columns=flexible,
basicstyle={\scriptsize\ttfamily},
numberstyle=\tiny\color{gray},
stringstyle=\color{mauve}
} 
\begin{lstlisting}[caption={Analyzer queries to find high priority broadcast blockers},label={lis:broadcastblockers},float=tb]
Q declaresEdges = universe.edgesTaggedWithAny(Edge.DECLARES).retainEdges();
Q callEdges = universe.edgesTaggedWithAny(Edge.CALL).retainEdges();
Q overridesEdges = universe.edgesTaggedWithAny(Edge.OVERRIDES).retainEdges();
Q abortBroadcast = methodSelect("BroadcastReceiver", "abortBroadcast").union(methodSelect("PendingResult", "abortBroadcast"));
abortBroadcast = abortBroadcast.union(overridesEdges.reverse(abortBroadcast));
Q onReceive = methodSelect("BroadcastReceiver", "onReceive");
onReceive = onReceive.union(overridesEdges.reverse(onReceive));
Q highPriorityTypes = context.nodesTaggedWithAny(AndroidManifest.MANIFEST_HIGH_PRIORITY.toString());
Q highPriorityOnReceive = onReceive.intersection(declaresEdges.forward(highPriorityTypes));
Q highPriorityBroadcastBlockers = callEdges.between(highPriorityOnReceive, abortBroadcast);
\end{lstlisting}

Listing~\ref{lis:broadcastblockers} shows the queries an analyzer could use to detect high priority broadcast blockers, which could be used to intercept and block SMS messages on an Android device.  Lines (1-3) select DECLARES, CALL, and OVERRIDES subgraphs; (4-5) selects abortBroadcast methods including overridden methods; (6-7) selects BroadcastReciever onReceive methods including overridden methods; (8) selects classes registered with a high priority in the Android manifest; (9) selects high priority onReceive methods; (10) selects CALL graphs that have an edge between the high priority onReceive methods and abortBroadcast methods.

\subsection{Dashboard}
\begin{figure}[t]
\begin{centering}
\includegraphics[width=0.43\textwidth]{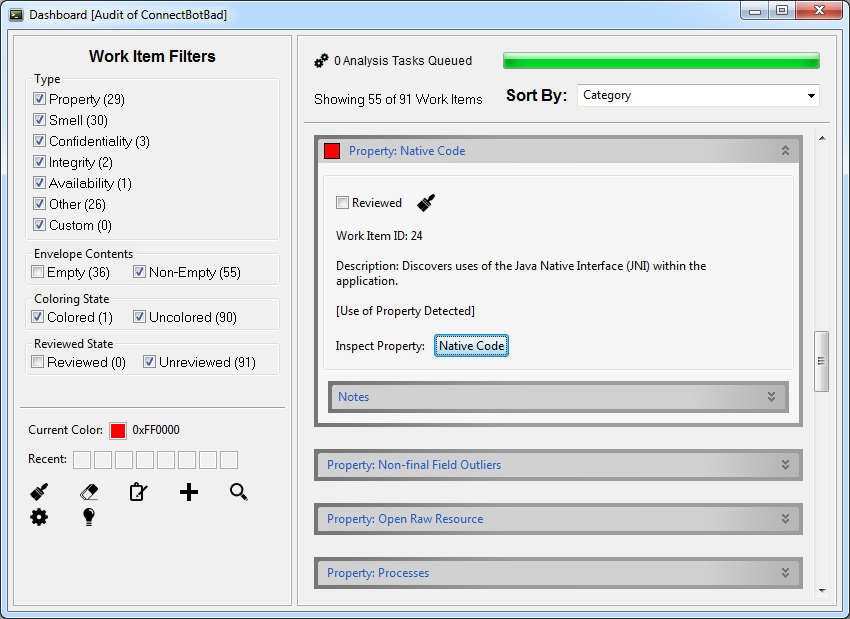}
\par\end{centering}

\caption{Security Toolbox Dashboard Interface}
\label{Fig:Security:Toolbox:Dashboard:Interface}
\end{figure}

The Dashboard (shown in Figure~\ref{Fig:Security:Toolbox:Dashboard:Interface}) is an interface for automating the execution and managing results of the Toolbox's automated analyzers.  The Dashboard accounts for analyzer dependencies to enable the highest amount of parallel computation while running a multitude of analyzers.  As results are computed, they are presented to the analyst in the work item queue on the right of the Dashboard.  Results can be filtered by category and marked as reviewed.  Optionally an analyst can make additional notes on a work item.  Since work items correspond to subgraphs of the program graph, they can be named and even colored to help identify separate program subsystems.  Program artifacts can be manually added or removed from a work item based on the colors given to program artifacts.

\subsection{Smart Views}
\label{sec:SmartViews}
Smart Views are developed from the observation that there are several graph traversal queries that analysts use over and over again during audits, such as forward and reverse control and data flows, or discovering the declarative structures and instantiations of an object.  To speed up such tasks, a graph for each of these queries can be automatically generated in response to mouse selection events on relevant source code or existing graph components.  Smart Views can be customized for particular Android-specific analysis tasks, such as showing user interface XML button event callbacks.

\begin{figure}[t]
\begin{centering}
\includegraphics[width=0.43\textwidth]{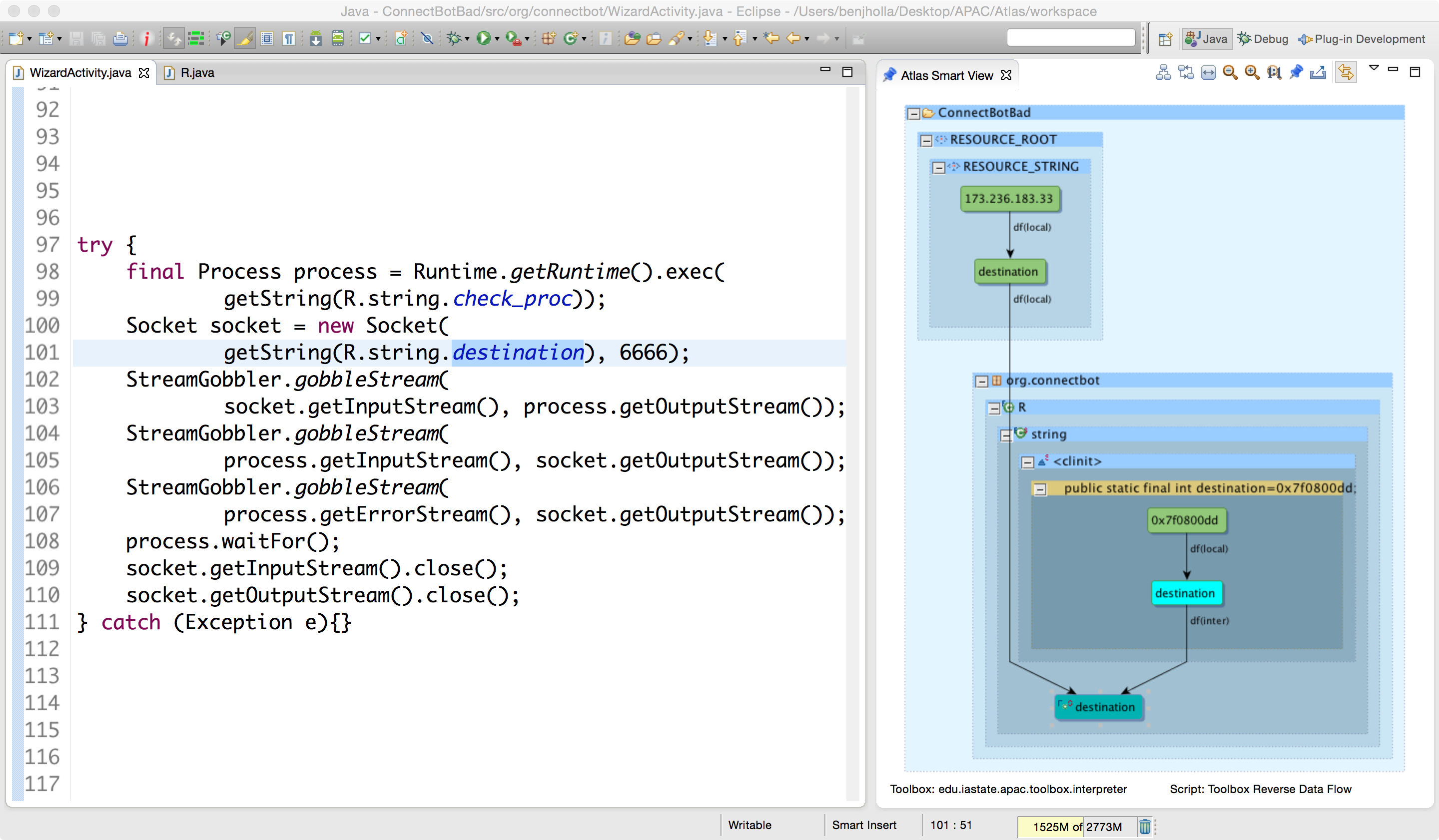}
\par\end{centering}

\caption{Smart View showing reverse data flow into Android XML resources from selected field}
\label{Fig:SmartView}
\end{figure}

Figure~\ref{Fig:SmartView} shows a customized Smart View showing a reverse data flow program slice that includes program artifacts in the Android XML resources.  The graph got generated when the user clicked on the "destination" field in the source window to inspect its value.
\section{Workflow}
\label{sec:Workflow}

The workflow of an audit follows a comprehension-driven model of an iterative Observe-Orient-Decide-Act (OODA) decision loop.  An audit starts by running the Dashboard, which produces evidence for the human analyst to inspect.  This information helps the analyst observe program behaviors and orient that information within the context of the application. Aided by this information, the analyst can prioritize his exploratory hypotheses to discover malice. To aid in testing a hypothesis, Smart Views are used to quickly follow control and data flows or perform a targeted analysis such as a symbolic analysis, Android Intent resolution, or matching exception throw and catch sites.  Confirming a hypothesis either results in the discovery of malware or results in more hypotheses to explore, which begins the process anew.  If a hypothesis set becomes depleted an audit is halted, and audits of remaining applications are reprioritized.  Finally, after the discovery of malware, the Security Toolbox is adapted by writing new analyzers to raise the bar for future automated analysis.
\balance
\section{Evaluation}
\label{sec:Evaluation}

By the end of Phase I of the DARPA APAC project, our team audited 77 Android applications developed by the Red team, of which 62 contained novel malware able to evade current automatic detection techniques.  DARPA employed a control team to use current state of the art tools to audit the apps along side Blue team performers.  Our process correctly classified 66 (85.7\%) apps as malicious or benign, found unintended malicious behaviours in 6 (7.8\%) apps, and missed malware in only 5 (6.5\%) of the apps consistently beating the control team.  We completed Phase I as the top performing Blue team.
\section{Related Work}
\label{sec:RelatedWork}

A number of tools and techniques have been developed for detecting malware in Android apps including some based on static analysis \cite{Fuchs2009, Payet2012, Wu2012,  zhou2012dissecting,fedler2013effectiveness, Aafer2013} and those based on dynamic analysis \cite{Enck2010, Reina2013}. These  automated  detection methods fall into two general categories: 1) signature-based and 2) machine learning-based. Signature-based approaches can be easily evaded by bytecode-level transformation attacks. Learning based
approaches extract features from application syntax, rather than program semantics, and are also subject to evasion.

Berkeley \cite{Felt:2011:APD:2046707.2046779} was the first to mine a mapping between Android permissions and the corresponding permission protected APIs using a dynamic analysis approach to randomly call APIs. Toronto later improved on Berkeley's incomplete mapping with a quicker, less involved, static analysis approach that mined complete public and private API mappings from the Android source code \cite{Au:2012:PAA:2382196.2382222}. Our Toolbox incorporates the Toronto mapping.

\section{Conclusion}
\label{sec:Conclusion}

Our novel human-in-loop approach to detect Android malware minimizes human effort by allowing the human to use the  evidence produced by the machine to focus their effort on further machine-assisted reasoning. This affords greater opportunity to detect malware that is not on the radar of an automated analyzer; the what-if experimentation capability provided by the machine enables the user to posit attacker intentions, hypothesize about the attacker's modus operandi and tailor queries to detect sophisticated malware. Thus, our approach increases automation, reduces human effort and error, and provides valuable machine assistance to detect novel and sophisticated malware.

This demo paper describes the Security Toolbox that implements our novel approach. The accompanying video shows a live audit that brings out various features of the Toolbox including the Dashboard (to run and manage automated analyzers), Permission Usage View (to list permissions and where they are used in the app), and Smart Views (to facilitate what-if experiments).  We acknowledge the valuable feedback from our reviewers of this paper.  Several components of the Security Toolbox are being open sourced under the MIT License at \url{https://github.com/EnSoftCorp}.

\bibliographystyle{IEEEtran}
\bibliography{IEEEabrv,references}

\begin{thebibliography}{10}
\providecommand{\url}[1]{#1}
\csname url@samestyle\endcsname
\providecommand{\newblock}{\relax}
\providecommand{\bibinfo}[2]{#2}
\providecommand{\BIBentrySTDinterwordspacing}{\spaceskip=0pt\relax}
\providecommand{\BIBentryALTinterwordstretchfactor}{4}
\providecommand{\BIBentryALTinterwordspacing}{\spaceskip=\fontdimen2\font plus
\BIBentryALTinterwordstretchfactor\fontdimen3\font minus
  \fontdimen4\font\relax}
\providecommand{\BIBforeignlanguage}[2]{{%
\expandafter\ifx\csname l@#1\endcsname\relax
\typeout{** WARNING: IEEEtran.bst: No hyphenation pattern has been}%
\typeout{** loaded for the language `#1'. Using the pattern for}%
\typeout{** the default language instead.}%
\else
\language=\csname l@#1\endcsname
\fi
#2}}
\providecommand{\BIBdecl}{\relax}
\BIBdecl

\bibitem{STUXNET}
R.~Langner, ``To kill a centrifuge,'' The Langner Group,
  http://www.langner.com/en/wp-content/uploads/2013/11/To-kill-a-centrifuge.pdf,
  Tech. Rep., nov 2013.

\bibitem{Heartbleed}
\BIBentryALTinterwordspacing
``Cve-2014-0160.'' [Online]. Available:
  \url{https://cve.mitre.org/cgi-bin/cvename.cgi?name=CVE-2014-0160}
\BIBentrySTDinterwordspacing

\bibitem{GOTO-Fail}
\BIBentryALTinterwordspacing
``Cve-2014-1266.'' [Online]. Available:
  \url{https://cve.mitre.org/cgi-bin/cvename.cgi?name=CVE-2014-1266}
\BIBentrySTDinterwordspacing

\bibitem{Shellshock}
\BIBentryALTinterwordspacing
``Cve-2014-6271.'' [Online]. Available:
  \url{https://cve.mitre.org/cgi-bin/cvename.cgi?name=CVE-2014-6271}
\BIBentrySTDinterwordspacing

\bibitem{Heartbleed-conspiracy}
\BIBentryALTinterwordspacing
``Heartbleed conspiracy theories.'' [Online]. Available:
  \url{http://www.businesscomputingworld.co.uk/openssl-heartbleed-criminal-and-government-conspiracy-theories/}
\BIBentrySTDinterwordspacing

\bibitem{Atlas-paper}
T.~Deering, S.~Kothari, J.~Sauceda, and J.~Mathews, ``Atlas: A new way to
  explore software, build analysis tools,'' in \emph{Companion Proceedings of
  the 36th International Conference on Software Engineering}, ser. ICSE
  Companion 2014.\hskip 1em plus 0.5em minus 0.4em\relax New York, NY, USA:
  ACM, 2014.

\bibitem{ICSE-Demo}
\BIBentryALTinterwordspacing
``Atlas video demo.'' [Online]. Available:
  \url{https://www.youtube.com/watch?v=cZOWlJ-IO0k}
\BIBentrySTDinterwordspacing

\bibitem{Au:2012:PAA:2382196.2382222}
\BIBentryALTinterwordspacing
K.~W.~Y. Au, Y.~F. Zhou, Z.~Huang, and D.~Lie, ``Pscout: Analyzing the android
  permission specification,'' in \emph{Proceedings of the 2012 ACM Conference
  on Computer and Communications Security}, ser. CCS '12.\hskip 1em plus 0.5em
  minus 0.4em\relax New York, NY, USA: ACM, 2012, pp. 217--228. [Online].
  Available: \url{http://doi.acm.org/10.1145/2382196.2382222}
\BIBentrySTDinterwordspacing

\bibitem{RTATypeAnalysis}
\BIBentryALTinterwordspacing
D.~F. Bacon and P.~F. Sweeney, ``Fast static analysis of c++ virtual function
  calls,'' in \emph{Proceedings of the 11th ACM SIGPLAN Conference on
  Object-oriented Programming, Systems, Languages, and Applications}, ser.
  OOPSLA '96.\hskip 1em plus 0.5em minus 0.4em\relax New York, NY, USA: ACM,
  1996, pp. 324--341. [Online]. Available:
  \url{http://doi.acm.org/10.1145/236337.236371}
\BIBentrySTDinterwordspacing

\bibitem{Fuchs2009}
A.~P. Fuchs, A.~Chaudhuri, and J.~S. Foster, ``{SCanDroid: Automated Security
  Certification of Android Applications},'' Department of Computer Science,
  University of Maryland, College Park, Tech. Rep., 2009.

\bibitem{Payet2012}
{\'E}.~Payet and F.~Spoto, ``Static analysis of android programs,''
  \emph{Information and Software Technology}, vol.~54, no.~11, pp. 1192--1201,
  2012.

\bibitem{Wu2012}
D.-J. Wu, C.-H. Mao, T.-E. Wei, H.-M. Lee, and K.-P. Wu, ``Droidmat: Android
  malware detection through manifest and api calls tracing,'' in
  \emph{Information Security (Asia JCIS), 2012 Seventh Asia Joint Conference
  on}.\hskip 1em plus 0.5em minus 0.4em\relax IEEE, 2012, pp. 62--69.

\bibitem{zhou2012dissecting}
Y.~Zhou and X.~Jiang, ``Dissecting android malware: Characterization and
  evolution.''\hskip 1em plus 0.5em minus 0.4em\relax IEEE, 2012.

\bibitem{fedler2013effectiveness}
R.~Fedler, J.~Schutte, and M.~Kulicke, ``On the effectiveness of malware
  protection on android,'' Fraunhofer AISEC, Tech. Rep., 2013.

\bibitem{Aafer2013}
Y.~Aafer, W.~Du, and H.~Yin, ``Droidapiminer: Mining api-level features for
  robust malware detection in android,'' in \emph{Proc. of International
  Conference on Security and Privacy in Communication Networks (SecureComm)},
  2013.

\bibitem{Enck2010}
W.~Enck, P.~Gilbert, B.-G. Chun, L.~P. Cox, J.~Jung, P.~McDaniel, and A.~Sheth,
  ``Taintdroid: An information-flow tracking system for realtime privacy
  monitoring on smartphones.'' in \emph{OSDI}, vol.~10, 2010.

\bibitem{Reina2013}
A.~Reina, A.~Fattori, and L.~Cavallaro, ``A system call-centric analysis and
  stimulation technique to automatically reconstruct android malware
  behaviors,'' in \emph{{Proceedings of the 6$^{th}$ European Workshop on
  System Security (EUROSEC)}}, {Prague, Czech Republic}, April 2013.

\bibitem{Felt:2011:APD:2046707.2046779}
\BIBentryALTinterwordspacing
A.~P. Felt, E.~Chin, S.~Hanna, D.~Song, and D.~Wagner, ``Android permissions
  demystified,'' in \emph{Proceedings of the 18th ACM Conference on Computer
  and Communications Security}, ser. CCS '11.\hskip 1em plus 0.5em minus
  0.4em\relax New York, NY, USA: ACM, 2011, pp. 627--638. [Online]. Available:
  \url{http://doi.acm.org/10.1145/2046707.2046779}
\BIBentrySTDinterwordspacing

\end{thebibliography}

\end{document}